# What Is the Dynamic Capacitance of a Memristor?

**Juan Bisquert* and Gonzalo Rivera-Sierra**
Instituto de Tecnología Química (Universitat Politècnica de València–Consejo Superior de Investigaciones Científicas), Av. dels Tarongers, 46022 València, Spain
Email: jbisquer@itq.upv.es

**Abstract**

Impedance spectroscopy of memristors, nanofluidic channels and halide perovskite devices often shows a low-frequency capacitance that is orders of magnitude larger than any plausible stored charge, or that becomes negative and appears together with an inductive loop. We distinguish two kinds of capacitance that enter the impedance in the same way: a storage capacitance, the derivative of a stored charge, and a dynamic capacitance, produced by the delayed response of a conduction current. Using the conductance-activated quasi-linear memristor (CALM) model, and a general memristor in which an internal state controls the conductance, we show that the dynamic capacitance is the time integral of the transient differential conductance after a small voltage step, measured from its final value. For a single relaxing state it is the product of a relaxing conductance and a relaxation time, with a negative sign. It vanishes at zero bias, and its sign is set by whether the stationary conductance increases or decreases with the magnitude of the voltage. Potentiation gives a negative capacitance and a chemical inductor, and depression gives a positive capacitance that is easily mistaken for stored charge. The analysis gives an explicit formulation of the time-domain criterion for negative capacitance of Jonscher and Ershov. In conical multipore nanofluidic membranes, the dynamic capacitance obtained from impedance and from voltage pulses shows the same sign change at zero bias and reaches values several hundred times larger than the membrane capacitance. The same picture accounts for the capacitive and inductive regimes of halide perovskite memristors.

## 1. Introduction: Storage and Dynamic Capacitance

In the elementary picture, a capacitance is the derivative of stored charge with respect to voltage. In impedance spectroscopy, however, capacitance is defined operationally from the admittance $Y(\omega)$ as $C(\omega) = \mathrm{Im}\, Y/\omega$. Many devices with slow internal dynamics, such as halide perovskite solar cells, memristors and nanofluidic channels, show measured capacitances that are far larger than any plausible stored charge, or that become negative [1–3]. In our early work on perovskite solar cells we found a giant low-frequency capacitance under illumination [1], and for a long time it was natural to look for the charge that it stored. The link between negative capacitance and conduction had been made much earlier. Jonscher, and later Ershov et al., related negative capacitance to a current that rises with time after a voltage step [4, 5]. Our own route to the same question came from the chemical inductor, and from the concepts of conduction capacitance and conduction inductance. These showed that a slow internal variable that modulates the current produces capacitive and inductive features without an additional reservoir of charge [6–8].

Two different kinds of capacitance therefore appear in impedance measurements. A storage capacitance is the derivative of a stored charge with respect to voltage. It includes the dielectric (geometric) and depletion capacitances, which describe electrostatic charge separation, the double-layer capacitance, and the chemical capacitance, which describes a change of carrier population [9, 10]. It exists also in equilibrium, without any current flowing through the device, and for stable passive storage it is positive. A dynamic capacitance, in contrast, is the contribution to the measured capacitance $\mathrm{Im}\, Y/\omega$ that arises from the delayed response of a conduction current. It requires a current flowing through the device and it can be positive or negative. It does not measure a stored charge, although the accumulation of some species may set its time scale. In a conical nanopore, for example, the conductance is controlled by the salt concentration near the tip, and the time needed to accumulate or deplete this salt sets the delay of the conductance [11, 12]. This accumulation does not separate charge across the device, so it does not appear as a storage capacitance. It enters the measured capacitance only through the delayed conductance it produces, that is, as a dynamic capacitance. Both kinds of capacitance enter the impedance spectrum in the same way, and an equivalent circuit alone cannot tell them apart. The low-frequency capacitance measured in a current-carrying device is in general the sum of the two.

Memristors provide the clearest setting to establish this distinction. Their defining property is a conductance controlled by an internal state that depends on the history of the applied voltage [13], and a delayed conductance is precisely the source of a dynamic capacitance. In neuromorphic computing, memristors emulate synapses, whose weight is the conductance. Trains of voltage pulses increase this weight, which is called potentiation, or decrease it, which is called depression, and in volatile devices the weight relaxes back when the stimulus stops, which provides a short-term memory [14, 15]. These functions are the basis of learning in hardware neural networks. In previous work we showed that potentiation has an inductive nature: when increasing the voltage raises the conductance, the current rises with a delay, which produces inverted hysteresis and an inductive impedance [6, 8, 14]. The dynamic capacitance is the small-signal counterpart of these functions. Its sign tells whether a device potentiates or depresses at a given bias, and its magnitude combines the amplitude of the conductance change with its memory time. It can therefore be read from a single impedance or transient measurement.

Here we first work out the dynamic capacitance in the CALM model, which we introduced for volatile and nonvolatile halide perovskite memristors [16]. We validated it in lead-free bismuth halide memristors [17] and also applied it to nanofluidic pores [18, 19]. In this model the storage capacitance is known by construction and all quantities can be calculated explicitly. We then show that the results hold for any memristor in which an internal state controls the conductance, and we use this picture to explain the capacitive and inductive responses of our nanofluidic and halide perovskite memristors.

## 2. The CALM Model

The CALM model is a particular case of a much broader class. Its structure, a conduction current set by a conductance, and a conductance set by internal variables that relax toward voltage-dependent stationary values, is that of the generic memristive systems of Chua [13]. It is also the structure of the conductance-based models of neurons, in which gating variables activate the ionic conductances of the membrane. Chua et al. showed that the ion channels of the Hodgkin–Huxley axon are memristors in this sense [20], and the linearization of these conductances had already revealed an apparent inductance and capacitance in the squid axon [21]. The same formalism describes volatile threshold switches whose state is the temperature of the active region, such as the Mott memristors used in neuristors [22], and compact models of resistive memories with a memory state [23]. It also covers the ionic devices discussed in this work: in nanofluidic channels the state is an ion concentration, and in halide perovskite memristors it is an ionic accumulation at an interface [19]. The results below therefore apply to a wide class of memristors and devices with memory, not only to the particular model.

Let $u$ be the voltage across the device and $x$ a normalized state variable between zero and one. The CALM equations, with a parallel geometric or membrane capacitance $C_m$, are [16]

$$I = I_c + C_m \frac{du}{dt}, \qquad I_c = G(x)\, u \tag{1}$$

$$G(x) = g_L + (g_H - g_L)x, \qquad g_H > g_L \tag{2}$$

$$\tau_k(u) \frac{dx}{dt} = x_{\mathrm{eq}}(u) - x \tag{3}$$

Here $g_L$ and $g_H$ are the low and high conductances, $x_{\mathrm{eq}}(u)$ is the state reached under a maintained voltage, and $\tau_k(u)$ is the relaxation time [15]. The model is quasi-linear because eqs 1–3 are linear in $x$, while the nonlinearity resides in the voltage dependence of $x_{\mathrm{eq}}$ and $\tau_k$. The two terms in eq 1 have different roles. The capacitor carries a charging current only while $u$ changes. The conduction term carries a sustained current under a constant voltage, and its magnitude evolves when $x$ changes. For the stationary state we use the sigmoid

$$x_{\mathrm{eq}}(u) = \frac{1}{1+\exp[-(u-V_B)/V_m]} \tag{4}$$

which increases monotonically with voltage for $V_m > 0$.

Consider now the device held at a constant (dc) bias $u_0$. The state has reached $x_0 = x_{\mathrm{eq}}(u_0)$, and the current has its stationary value. A small perturbation $\delta u$ superimposed on this bias produces deviations $\delta x$ and $\delta I$ from these stationary values. From eq 1, the conduction current changes both directly with $\delta u$ and through the change of state $\delta x$. It is convenient to define three local quantities: the instantaneous conductance $g_b = G(x_0)$, the slope of the stationary state $a = dx_{\mathrm{eq}}/du$ at $u_0$, and the time $\tau_0 = \tau_k(u_0)$. The coupling of the state to the current is then described by the relaxing conductance $g_a$, and the response after full adaptation by the total conductance $g_{\text{tot}}$:

$$g_a = u_0(g_H - g_L)a, \qquad g_{\text{tot}} = \frac{dI_{\text{ss}}}{du} = g_b + g_a \tag{5}$$

Here $g_{\text{tot}}$ is the differential conductance of the stationary current–voltage curve $I_{\text{ss}}(u)$. The three conductances are local quantities at the bias $u_0$, and should not be confused with the function $G(x)$. The instantaneous conductance $g_b$ is the response when the state is frozen; $g_{\text{tot}}$ is the response when the state has fully adapted; and $g_a$ is the difference between the two. The voltage dependence of $\tau_k$ does not enter the linear response. When eq 3 is linearized, the derivative $d\tau_k/du$ multiplies $dx/dt$, which vanishes at the stationary point, while $\tau_0$ remains as the coefficient of the time derivative of $\delta x$. Only the value $\tau_0$ matters, not its slope with voltage.

## 3. Dynamic Capacitance in the Frequency and Time Domains

For a small perturbation of angular frequency $\omega$, $\delta u \propto \exp(i\omega t)$, the linearized state equation gives the response of the state $\delta x$, and the admittance $Y = \delta I/\delta u$ is

$$\delta x = \frac{a\,\delta u}{1+i\omega\tau_0}, \qquad Y(\omega) = g_b + \frac{g_a}{1+i\omega\tau_0} + i\omega C_m \tag{6}$$

Separating real and imaginary parts,

$$\operatorname{Re} Y = g_b + \frac{g_a}{1+(\omega\tau_0)^2}, \qquad \operatorname{Im} Y = \omega C_m - \frac{g_a\,\omega\tau_0}{1+(\omega\tau_0)^2} \tag{7}$$

and the measured capacitance is

$$C(\omega) = \frac{\operatorname{Im} Y}{\omega} = C_m - \frac{g_a\tau_0}{1+(\omega\tau_0)^2} \tag{8}$$

The capacitance contains two contributions of different nature. The first, $C_m$, is the storage capacitance of the device. The second is the dynamic capacitance produced by the relaxation of conduction, whose low-frequency value is

$$C_{\text{dyn}} = -g_a\tau_0 \tag{9}$$

It persists in the limit $\omega \to 0$, although the phase of the conduction current vanishes, because the capacitance is obtained by dividing $\operatorname{Im} Y$ by $\omega$. At $\omega\tau_0 \gg 1$ the state cannot follow the perturbation, and only $C_m$ remains. Equation 7 shows that the real part changes from $g_{\text{tot}}$ at low frequency to $g_b$ at high frequency, so that $g_a$ is read directly from the difference of the two plateaus. The dependence of the dynamic capacitance on bias and frequency is illustrated in Figure 1b,d.

The admittance also determines the response in the time domain. By a general result of linear response [5], the admittance is $s$ times the Laplace transform of the current response to a small voltage step $\Delta u$ applied at $t = 0$ to a device at a stable stationary point,

$$Y(s) = s\int_0^\infty e^{-st}\,\frac{\delta I(t)}{\Delta u}\,dt, \qquad s = i\omega \tag{10}$$

where $\delta I$ is the total current, the sum of the capacitive current and the conduction current of eq 1. For the admittance of eq 6, the capacitive current is a short initial spike that delivers the charge $C_m\Delta u$. The change of the conduction current divided by the step amplitude defines a transient differential conductance:

$$\frac{\delta I_c(t)}{\Delta u} = g_b + g_a[1 - \exp(-t/\tau_0)] \quad (11)$$

Immediately after the step the state has not changed, and the conductance is $g_b$; as the state relaxes, it reaches the final value $g_{\text{tot}}$. If $g_a > 0$ the current rises toward its final value, as in a delayed activation of a conducting channel. If $g_a < 0$ the current first overshoots and then decays.

To obtain the low-frequency limit, we write the total response $\delta I(t)/\Delta u$ as its final value $g_{\text{tot}}$ plus a deviation $y(t)$ that vanishes at long times. Since the Laplace transform of the constant $g_{\text{tot}}$ is $g_{\text{tot}}/s$, eq 10 becomes

$$Y(s) = g_{\text{tot}} + s\int_0^\infty e^{-st}\, y(t)\, dt \;\approx\; g_{\text{tot}} + i\omega \int_0^\infty y\,(t)\, dt \quad (12)$$

where the approximation holds when $\omega$ is small compared with the inverse of all the relaxation times, so that $e^{-st} \approx 1$ wherever $y(t)$ is not negligible. The coefficient of $i\omega$ is the low-frequency capacitance. Therefore, provided the transient is integrable, the low-frequency capacitance is the area of the current transient relative to its final value:

$$C_{\text{LF}} = \frac{1}{\Delta u}\int_0^\infty [\delta I(t) - \delta I(\infty)]\, dt = C_m + C_{\text{dyn}} \quad (13)$$

The charging spike of the capacitor contributes the storage capacitance $C_m$. The conduction transient of eq 11 contributes the dynamic capacitance

$$C_{\text{dyn}} = \int_0^\infty \left[\frac{\delta I_c(t)}{\Delta u} - g_{\text{tot}}\right] dt = -g_a \int_0^\infty e^{-t/\tau_0}\, dt = -g_a \tau_0 \quad (14)$$

Equation 14 expresses the dynamic capacitance as the time integral of the transient differential conductance, measured from its final value. It is the charge that the conduction current delivers in excess of, or in deficit to, its final stationary flow, divided by the step amplitude. A rising current gives a negative area and a negative capacitance; an overshoot gives a positive one. The dynamic term can easily dominate, because a relaxing conductance multiplied by a slow time can be many times larger than the membrane capacitance.

In the CALM model, around a stationary state, the sign of $g_a$ is the sign of the bias, since $g_H > g_L$ and the slope $a$ of the sigmoid of eq 4 is positive. At positive bias an increase of the voltage raises the conductance and the current, $g_a > 0$, and the response is inductive. At negative bias an increase of the voltage magnitude lowers the conductance, $g_a < 0$, and the response is capacitive. At $u_0 = 0$ the coupling vanishes, and the dynamic capacitance with it, while the storage capacitance remains finite. The same kinetic law therefore gives both signs of capacitance, and only the operating point changes. These results refer to small perturbations around a stationary state, which is what impedance spectroscopy and long small-step transients measure, and we keep this condition throughout this work. Under trains of short pulses the state does not reach its stationary value, the response depends on the history of the state, and its sign can differ; we return to this regime in the Outlook.

Figure 1 illustrates these results with the activation and relaxation-time parameters that we determined for a conical multipore nanofluidic membrane, discussed in Section 7 [19]. The relaxing conductance changes sign at zero bias (Figure 1a), and the low-frequency dynamic capacitance $-g_a\tau_0$ changes sign with it (Figure 1b). At $-0.4$ V we obtain $g_b = 1.52$ µS, $g_a = -0.386$ µS and $\tau_0 = 1.13$ ms, hence $C_{\text{dyn}} = +0.44$ nF. At $+0.4$ V, $g_b = 3.05$ µS, $g_a = +1.24$ µS and $\tau_0 = 9.03$ ms, hence $C_{\text{dyn}} = -11.2$ nF. The corresponding step transients (Figure 1c) show the decaying overshoot at negative bias and the delayed rise at positive bias, and the shaded areas are the capacitances of eq 14. Figure 1d shows the frequency dependence of eq 8: the dynamic capacitance is constant at low frequency, has decreased to half its value at $\omega\tau_0 = 1$, and vanishes at high frequency.

We used the same analysis of small-step transients to map the internal ionic and electronic currents of halide perovskite solar cells [24, 25]. There, the delayed rise of the ion-controlled recombination current, whose area is the negative dynamic capacitance of eq 14, coexists with the decaying charging current of an interfacial storage capacitance, so that both kinds of capacitance appear in the same transient.

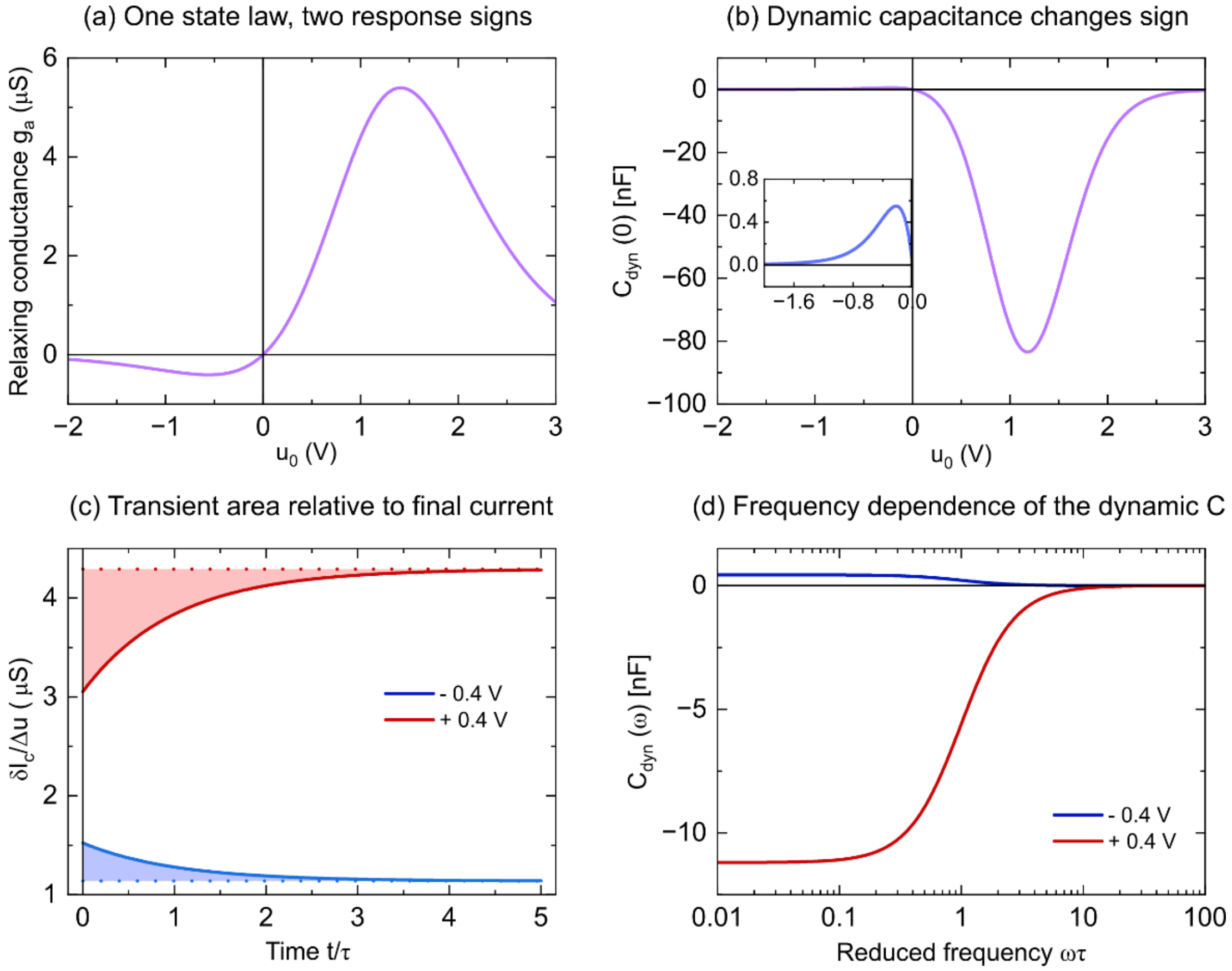

Figure 1. Calculated CALM response. (a) Relaxing conductance $g_a$ as a function of bias. (b) Low-frequency dynamic capacitance $-g_a\tau_0$; the inset shows the positive values at negative bias. (c) Conduction current after a small positive voltage step at $-0.4$ V and $+0.4$ V, normalized by the step. Dashed lines mark the final values, and the shaded areas give the dynamic capacitance. (d) Dynamic capacitance as a function of reduced frequency $\omega\tau_0$. The charging current of $C_m$ is not included in (c) and (d). Parameters: $V_B = 1.023$ V, $V_m = 0.511$ V and $\tau_k = Ka^{\alpha} + t_0$, with $K = 0.064$ s V$^{\alpha}$, $\alpha = 1.849$ and $t_0 = 0.104$ ms, from ref 19; $g_L = 1$ μS and $g_H = 10$ μS are illustrative.

## 4. Equivalent Circuits and Hysteresis

The admittance of eq 6 corresponds to the equivalent circuits of Figure 2. For $g_a > 0$, the relaxing term is the admittance of a resistance in series with an inductance (Figure 2a),

$$\frac{g_a}{1+i\omega\tau_0} = \frac{1}{R_a+i\omega L_a}, \qquad R_a = \frac{1}{g_a}, \qquad L_a = \frac{\tau_0}{g_a} \tag{15}$$

This is the chemical inductor [6]: the inductance expresses the delay of the conductance activation and has no magnetic origin. For $g_a < 0$, writing $A = -g_a > 0$ and rearranging eq 6 gives

$$g_b - \frac{A}{1+i\omega\tau_0} = g_{\text{tot}} + \frac{i\omega C_d}{1+i\omega\tau_0}, \qquad C_d = -g_a\tau_0, \qquad R_d = \frac{1}{A} \tag{16}$$

For $g_{tot} > 0$, this is an ordinary resistance $1/g_{tot}$ in parallel with a series $R_d C_d$ branch with positive capacitance (Figure 2b). The two circuits describe the same physical process at opposite polarity. An important practical consequence is that the RC branch of Figure 2b has the same form as a double-layer capacitance charged through a resistance. A positive capacitive arc therefore does not identify by itself the origin of the response. What distinguishes the conduction mechanism is its voltage dependence: the branch changes sign with bias, and its amplitude $g_a$ is fixed by the stationary current–voltage curve.

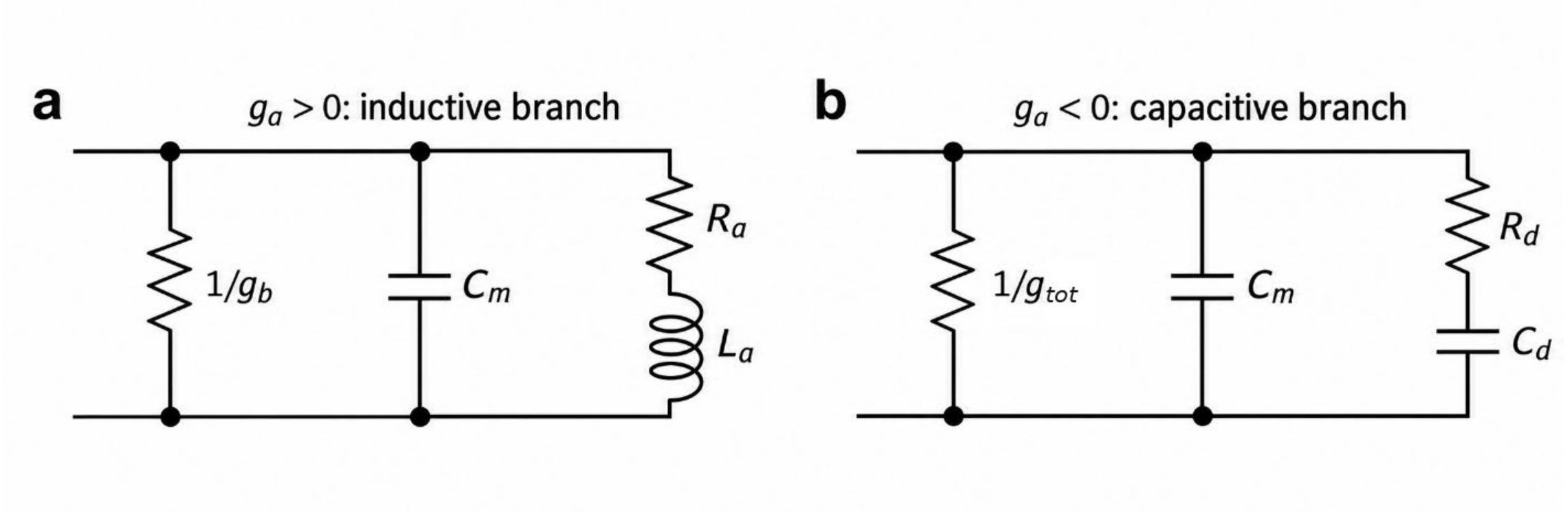

Figure 2. Equivalent circuits of the CALM admittance, eq 6. (a) For positive relaxing conductance, $g_a > 0$, the slow branch is a resistance $R_a = 1/g_a$ in series with an inductance $L_a = \tau_0/g_a$. (b) For $g_a < 0$, the same admittance is a resistance $1/g_{tot}$ in parallel with a series branch with $R_d = -1/g_a$ and positive capacitance $C_d = -g_a\tau_0$. In both cases $C_m$ is the membrane capacitance.

The same coefficient controls the current–voltage hysteresis. In a slow voltage sweep the state lags slightly behind its stationary value, by an amount proportional to the scan rate, with $a$ and $\tau_0$ evaluated at the instantaneous voltage. Substitution in eq 1 gives the current:

$$x \approx x_{eq}(u) - a\,\tau_0 \frac{du}{dt}, \qquad I \simeq I_{ss}(u) + (C_m - g_a\tau_0)\frac{du}{dt} \tag{17}$$

A loop in which the forward scan, toward larger voltage magnitude, carries more current than the reverse scan is called regular, or capacitive, hysteresis. This is the response of a capacitor, and in the first quadrant the loop runs clockwise. In the opposite case, called inverted, or inductive, hysteresis, the reverse scan carries more current and the loop runs counterclockwise. These terms are widely used for perovskite solar cells [8]. Equation 17 shows that a positive low-frequency capacitance produces regular hysteresis and a negative one produces inverted hysteresis [8]. This relation connects impedance spectra and current–voltage curves through a single parameter, although large switching loops require integration of the full nonlinear equations [16].

## 5. Memristors with a State-Controlled Conductance

The results obtained with the CALM model can be formulated more generally. Consider a general voltage-controlled memristor in which an internal state $x$ controls the conductance [13]:

$$I = C_m \frac{du}{dt} + G(x)\,u, \qquad \tau_k \frac{dx}{dt} = f(u, x) \tag{18}$$

where $\tau_k > 0$ is a kinetic time constant and $f$ is a rate function. A stationary point $x_0$ at the bias $u_0$ satisfies $f(u_0, x_0) = 0$, and it is stable if $\partial f/\partial x < 0$. The linearization of eq 18 gives an admittance of the same form as eq 6 [6, 7], with

$$g_b = G(x_0), \qquad g_a = \frac{u_0 \frac{dG}{dx}\frac{\partial f}{\partial u}}{-\frac{\partial f}{\partial x}}, \qquad \tau_0 = \frac{\tau_k}{-\partial f/\partial x} \tag{19}$$

so that the dynamic capacitance is

$$C_{\text{dyn}} = -g_a \tau_0 = -\frac{u_0 \tau_k}{(\partial f/\partial x)^2}\frac{dG}{dx}\frac{\partial f}{\partial u} \tag{20}$$

For the CALM model, $dG/dx = g_H - g_L$, $\partial f/\partial u = a$ and $\partial f/\partial x = -1$, which recovers eq 5. Two general properties follow from eq 20 for regular functions $G$ and $f$. First, as in the CALM model, the dynamic capacitance vanishes at zero bias. Second, its sign is fixed by the sign of the product $u_0\,(dG/dx)(\partial f/\partial u)$. By eq 19, $g_a = u_0\, dG_{\text{ss}}/du$, where $G_{\text{ss}}$ is the conductance in the stationary state. It is therefore positive when an increase of the voltage magnitude raises the stationary conductance, and negative when it lowers it. These are the small-signal counterparts of potentiation and depression [14], and we use these terms in this sense below. Under potentiation the transient differential conductance rises toward its final value, and the dynamic capacitance is negative, that is, inductive. Under depression it decays after an overshoot, and the dynamic capacitance is positive. For a monotonic bipolar activation, as in the CALM model and in conical nanopores, the dynamic capacitance changes sign with the polarity. If instead the state responds to the magnitude of the voltage and raises the conductance as the magnitude increases, the product is positive at both polarities. The response is then inductive on both sides of zero bias, a dual-inductive behavior that we observed in halide perovskite memristors with ion-modulated interfacial barriers [26].

Equations 10–14 coincide with the small-signal relations between transient current and capacitance established by Jonscher and by Ershov et al. [4, 5]. In their work, negative capacitance results from a current that rises after a voltage step, and the low-frequency capacitance equals the geometric capacitance plus the net area of the relaxation current, as in eq 13. Ershov et al. also noted that the incremental charge method, $C = \Delta Q/\Delta u$, fails when a conduction current flows [5]. Our analysis gives this picture an explicit formulation in terms of state kinetics and conductance, with consistent frequency- and time-domain expressions, in three respects.

First, it expresses the cause of the rising current as an internal state that raises the conductance with a delay. The mechanisms proposed in those works belong to this class, and eq 20 gives the general condition. They include the relief of space charge by carriers of opposite sign [4] and injection controlled by the recharging of quantum wells [5].

Second, it treats both signs on the same footing: depression gives a positive area that is equally a conduction effect, so that a positive low-frequency capacitance can also originate in conduction rather than in stored charge.

Third, it explains the polarity dependence observed by Jonscher in Schottky diodes, with negative capacitance under forward bias but not under reverse bias [4]. This is the behavior expected from eq 20 for a monotonic bipolar activation.

Finally, Jonscher discarded an inductive interpretation because an inductance seemed to require magnetic energy storage [4]. The chemical inductor of eq 15 removes this objection. At $\omega\tau_0 \gg 1$ its branch gives $C \approx -1/(\omega^2 L_a)$, the frequency dependence expected for an inductor, but it originates in the delayed activation of conduction and stores no magnetic energy [6].

## 6. Measuring the Dynamic Capacitance

The two parameters that determine the dynamic capacitance, the relaxing conductance $g_a$ and the relaxation time $\tau_0$, can be measured independently at a stationary bias. There are three methods. They rely on different measurements, so that each one provides a check of the others.

*Impedance spectroscopy.* The real part of the admittance (eq 7) shows two plateaus, $g_{\mathrm{tot}}$ at low frequency and $g_b$ at high frequency. Their difference gives $g_a$. The transition between the plateaus occurs at the characteristic frequency $\omega = 1/\tau_0$, which gives the relaxation time. Equivalently, both parameters are obtained by fitting the spectrum with one of the equivalent circuits of Figure 2.

*Small-step transients.* After a small voltage step $\Delta u$ at the same bias, the conduction current changes from its initial value $g_b \Delta u$ to its final value $g_{\mathrm{tot}} \Delta u$, with the time constant $\tau_0$ (eq 11). The amplitude of the change gives $g_a$. In addition, the area of the transient relative to its final value gives the dynamic capacitance directly (eq 14), without assuming a particular form of the relaxation.

*Stationary current–voltage curve.* In the CALM model the stationary current is $I_{\mathrm{ss}} = G_{\mathrm{ss}} u$, with $G_{\mathrm{ss}} = G(x_{\mathrm{eq}})$. The amplitude can therefore also be obtained from a slow, steady-state measurement:

$$g_a = \frac{dI_{\mathrm{ss}}}{du} - \frac{I_{\mathrm{ss}}}{u} = u \frac{dG_{\mathrm{ss}}}{du} \qquad (21)$$

This method does not give $\tau_0$, but it provides an independent check of the amplitude, because it uses only stationary data.

Agreement between these estimates supports the interpretation of the measured response as a relaxation of conduction within the assumed model. In Section 7.1 we compare the first two methods in a conical multipore membrane. There the transients are obtained with large voltage pulses from zero bias rather than with small steps, and their analysis requires the extension of eq 14 to large steps, which we give in the Supporting Information (eq S1).

Table 1 summarizes the experimental signatures that distinguish the two kinds of capacitance.

Table 1. Experimental Signatures of Storage and Dynamic Capacitance

| **Signature** | **Storage capacitance** | **Dynamic capacitance** |
|---|---|---|
| Origin | Stored charge (dielectric, double layer, chemical) | Delayed response of the conduction current |
| Requires current | No | Yes |
| Sign | Positive | Either sign |
| Zero bias ($I = G(x)u$) | Finite | Vanishes |
| Small-step transient | Charging spike | Rising current or overshoot |
| Equivalent circuit | Capacitor | Chemical inductor ($g_a > 0$) or RC branch ($g_a < 0$) |
| Current–voltage hysteresis | Regular | Regular or inverted, set by $-g_a \tau_0$ |
| Consistency test | Independent of $g_a$ | $g_a$ from impedance equals $u\, dG_{\mathrm{ss}}/du$ (eq 21) |

## 7. Nanofluidic Memristors

### 7.1. Impedance and Pulses in Conical Multipore Membranes

The membranes of our study contain about 300 conical nanopores with negatively charged walls [19]. Under positive bias the mobile-ion concentration near the pore tips increases and the conductance rises, while negative bias depletes this region and lowers the conductance. The state variable $x$ is therefore a normalized ion concentration. The impedance spectra (Figure 3a) show capacitive arcs at negative bias and a low-frequency inductive loop at positive bias. In the fitted circuit (Figure S1a), the relaxing conductance changes sign from negative to positive across zero bias, whereas $g_b$ remains positive (Figure S1b), as predicted by eq 5. The membrane capacitance is about 0.2 nF and nearly independent of bias (Figure S1c). The fitted inductance $L_a = \tau_0/g_a$ carries the sign of $g_a$ (Figure S1d), and its negative values at negative bias simply express the positive capacitance of Figure 2b in a single formula. At high positive bias the fit requires an additional ionic RC branch, which is a separate process and is not part of the CALM relaxation (Figure S1e,f).

Figure 3 shows the central experimental result of this work. From the fitted parameters we obtain the dynamic capacitance as $C_d = -g_a\tau_0 = -g_a^2 L_a$. Independently, we applied voltage pulses of increasing amplitude, each starting from 0 V, at both polarities (Figure S2), and integrated the current relative to its final value, as in eq 14. At negative bias the magnitude of the current decays after the step, which gives a positive area (Figure 3b). At positive bias the current rises toward its final value, which gives a negative area (Figure 3c). The two estimates show the same behavior (Figure 3d). The dynamic capacitance is small and positive at negative bias, below 1 nF (inset of Figure 3d). It changes sign at zero bias and becomes large and negative at positive bias. Its minimum is about $-100$ nF near 1 V in the impedance data and about $-70$ nF near 2 V in the pulse data. Both values are several hundred times larger than the membrane capacitance. The same kinetics explains the response to longer pulse sequences, in which trains of positive and negative pulses produce conductance potentiation and depression [19].

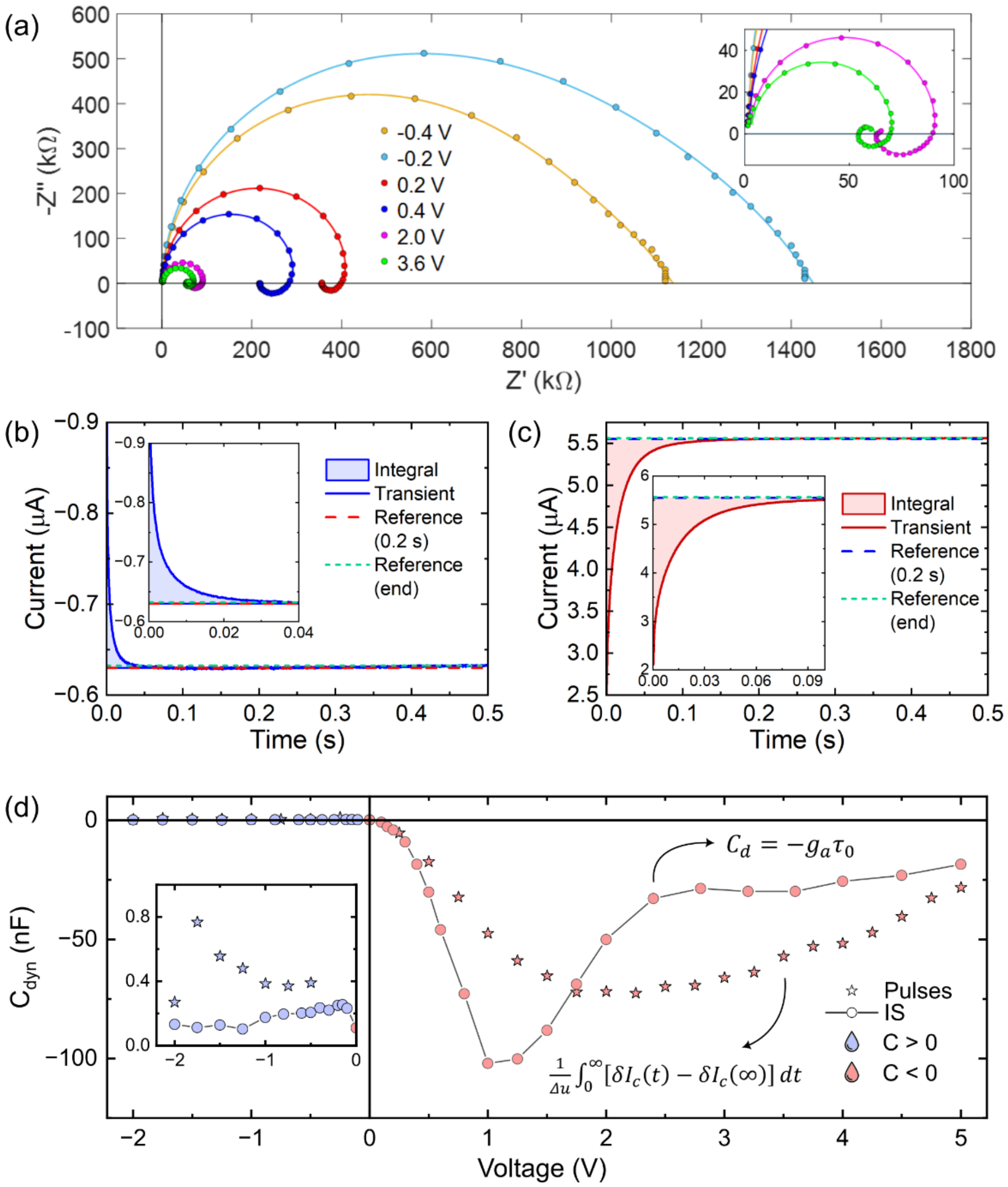

Figure 3. Dynamic capacitance of a conical multipore nanofluidic membrane from impedance and from voltage pulses. (a) Impedance spectra and fits at several biases; the inset shows the high-bias spectra. Reproduced from ref 19 under CC BY 4.0. (b) Current transient after a negative voltage step and (c) after a positive voltage step; the insets show the first part of the transient. The shaded areas are the integrals of eq 14, taken relative to the current at 0.2 s or at the end of the pulse (dashed lines). (d) Dynamic capacitance as a function of voltage, from the fitted impedance parameters, $C_d = -g_a\tau_0$ (circles), and from the transient areas (stars). Blue and red symbols indicate positive and negative values; the inset shows the positive values at negative bias. The fitted parameters are shown in Figure S1 and the complete pulse sequences in Figure S2.

The two estimates agree in sign and order of magnitude, but not exactly. This is expected for two reasons. First, the pulses are large steps from zero bias, so that their area measures the average slope of the stationary conductance between 0 and $V$ rather than its local slope at $V$ (Section S3). Second, the transients contain all the processes of the device, including the slow ionic branch that appears at high positive bias (Section S4 and Figure S3). The impedance fit separates this branch, and therefore gives the more direct estimate of $C_{\mathrm{dyn}}$, while the pulses confirm its sign and magnitude in the time domain. The pulse transients and their comparison with the impedance results are discussed in detail in the Supporting Information (Sections S2–S4).

### 7.2. Other Nanofluidic Systems

The coexistence of a capacitive response at one polarity and an inductive response at the other is not specific to our membranes. It is a general and characteristic observation in nanofluidic channels. Membranes of conical nanopores show capacitive arcs at reverse bias and inductive features at forward bias [27, 28], and series and parallel arrangements of such membranes combine both responses in a predictable way [28]. In single conical nanopipettes, the current decays after a voltage step of one sign and rises after a step of the opposite sign, an effect described as a negative memory capacitance [29]. In nanochannels formed by stacked gold nanoparticles, the response changes from capacitive to inductive when the electrolyte concentration increases [30]. These observations are usually explained by a different mechanism for each sign, such as double-layer charging for the capacitive arc and ion pairing or a chemical inductor for the inductive one. Our analysis suggests a simpler reading: both responses can be the two signs of a single dynamic capacitance $-g_a\tau_0$. As noted for Figure 2b, a large low-frequency capacitance, such as the value of about 185 μF reported in ref 30, should not be assigned to stored charge before a conduction contribution has been examined. This can be done through its change of sign with the operating conditions and through eq 21.

## 8. Halide Perovskite Memristors

Halide perovskite memristors show how the dynamic capacitance depends on the activation of the conductance and on the retention of the state, across different switching mechanisms. In early work, we observed an inductive arc emerging during the transition to the low-resistance state [31]. Figure 4 brings together three examples from our studies: volatile activation at one polarity, nonvolatile bipolar switching, and volatile activation at both polarities [17, 26, 32].

In the volatile $MAPbBr_3$ device, increasing the positive voltage produces inductive current–voltage hysteresis, and reducing it returns the conductance to the OFF state (Figure 4a) [32]. The impedance changes from capacitive at +0.2 V to inductive at +1.0 V (Figure 4b,c). The current decays during pulses of low amplitude, and it shows a delayed rise at higher amplitude, which becomes prominent during longer pulses (Figure 4d,e). This delayed rise is the time-domain form of the chemical inductor, and it appears in the voltage and time ranges where potentiation occurs. Within a monotonic CALM activation, $g_a$ is positive at any positive stationary bias, so that the dynamic contribution is negative even where the measured capacitance is positive. At low bias the activation is weak and $|g_a\tau_0|$ is small, and the storage capacitance of the interfacial ions dominates the low-frequency response until the dynamic contribution becomes larger [33, 34].

The nonvolatile $Cs_3Bi_2I_6Br_3$ device represents a different regime, in which switching is attributed to the formation and rupture of conductive filaments [17]. Its bipolar hysteresis and the retention of its resistance states are shown in Figure 4f,g. That study did not report impedance or small-step transients, but the fitted CALM parameters allow the stationary dynamic capacitance to be predicted. We calculate $g_a$ from the derivative of the sigmoid stationary conductance and obtain $C_{\mathrm{dyn}} = -g_a\tau_0$ in Figure 4h. The prediction shows the same bipolar sign change as the nanopore model, positive at negative bias and negative at positive bias. The long relaxation times of the fit near zero bias make both contributions very large, about +21 and $-155$ mF. These values are the very-low-frequency limit around a stationary state, not the capacitance during a fast writing pulse, and their magnitude is a model extrapolation that requires verification by impedance or small-step transients.

Perovskite memristors with ion-modulated interfacial barriers show dual-inductive switching (Figure 4i,j) [26]. Both polarities activate a volatile conducting state, which gives two complementary inductive lobes in one device. This behavior does not require two independent memory variables. In the terms of Section 5, a stationary conductance that increases with $|u|$ gives $g_a > 0$ on both sides of zero bias, so that the dynamic capacitance is negative in both activated regions and vanishes at $u = 0$.

Related devices show a chemical inductor accompanying nonvolatile potentiation under long pulses [35], and lateral $MAPbI_3$ memristors exhibit decaying and rising transients under different operating conditions [36]. These examples show that delayed conduction is common to different switching mechanisms and retention times, and that in each case its contribution must be separated from storage.

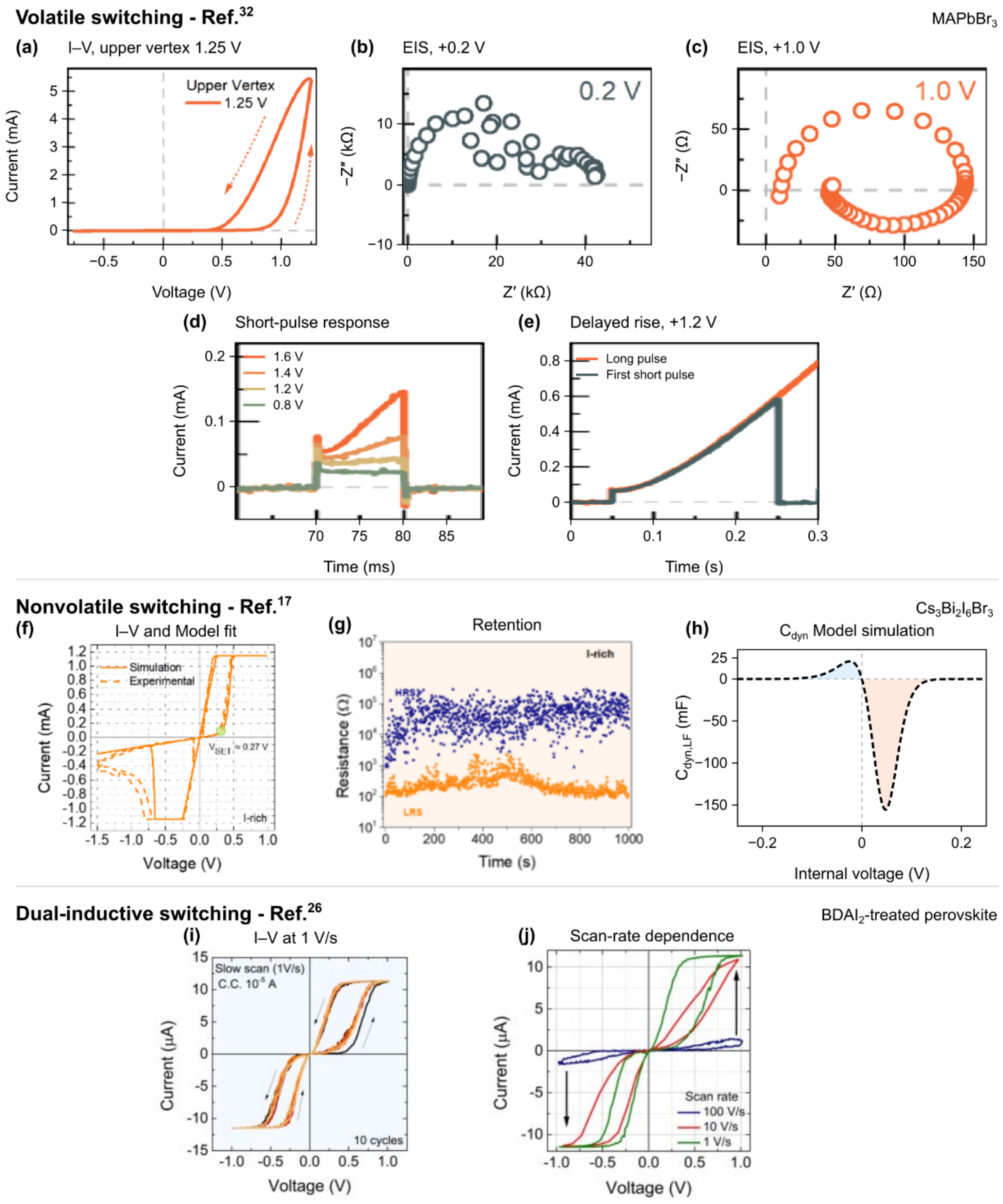

Figure 4. Three switching regimes in halide perovskite memristors. Volatile $MAPbBr_3$: (a)

current–voltage hysteresis at 1 V $s^{-1}$ (upper vertex 1.25 V); (b,c) impedance at +0.2 and +1.0 V; (d) responses to 10 ms pulses; and (e) initial response at +1.2 V to a 5 s pulse and to a 200 ms pulse (train period 250 ms). Nonvolatile $Cs_3Bi_2I_6Br_3$: (f) experimental and CALM-simulated hysteresis at 10 V $s^{-1}$; (g) retention at +0.10 V; and (h) dynamic capacitance predicted from the CALM parameters of ref 17. Dual-inductive switching: (i) ten cycles at 1 V $s^{-1}$ with 10 μA compliance, and (j) scan-rate dependence. Panels (a–e) adapted from ref 32, © 2024 The Authors, American Chemical Society. Panels (f,g) adapted from ref 17 and its Supporting Information. Panels (i,j) adapted from ref 26, © 2026 The Authors, Wiley-VCH GmbH. Adapted material is licensed under CC BY 4.0. Panel (h) is calculated here.

## 9. Related Effects in Perovskite Solar Cells

The distinction between storage and dynamic capacitance also applies to perovskite solar cells. Their conduction law need not have the form $I = G(x)u$, so that the rules for zero bias and polarity derived above are not general properties of photovoltaic devices, but the relation between the transient area and the admittance remains valid. Jacobs et al. separated charge storage from ion-modulated recombination, whose contribution to the measured capacitance can be large and of either sign [2]. Moia et al. described how the redistribution of ions controls interfacial injection and recombination, amplifying a small ionic current into a larger electronic one [37], and Ebadi et al. related the apparent light-enhanced and negative capacitances to slow current transients [38]. In our $MAPbBr_3$ solar cells, a capacitive arc evolves into a chemical inductor as the bias increases, and the continuity of the characteristic time indicates a common ion-controlled process [39]. In these devices the interfacial ions both store charge and control the electronic current, so that the low-frequency capacitance contains a storage and a dynamic contribution, and the dynamic part can be negative or positive depending on whether the ions reinforce or oppose the electronic current. We will present a detailed analysis of perovskite solar cells in these terms separately.

## 10. Outlook

The distinction between storage and dynamic capacitance opens several questions that we consider important for the field.

**From small signal to pulse trains.** The dynamic capacitance is a small-signal quantity, while synaptic functions are tested with trains of large pulses. The large-step result of eq S1 is a first bridge between the two: in the CALM model, the area of a single large step is set by the same relaxation time and by the change of stationary conductance between the initial and final voltages. Pulse trains that do not reach the stationary state, and the nonvolatile component of potentiation observed in perovskite memristors [35], require the full nonlinear dynamics. An open question is how far the impedance measured at a stationary bias predicts the change of synaptic weight under a given pulse protocol.

**Engineering the sign.** The sign of the dynamic capacitance is set by how the state depends on the voltage. Monotonic bipolar activation gives opposite signs at the two polarities, while activation by the voltage magnitude gives an inductive response at both [26], as do suitable combinations of nanopores [28]. The sign and magnitude of $-g_a\tau_0$ thus become design parameters for synapses, filters and inductive elements in neuromorphic circuits.

**Storage and conduction together.** Real devices contain both kinds of capacitance, and their competition produces characteristic features. When the conduction law has the form $I = G(x)\,u$, the dynamic capacitance vanishes at zero bias, so that the opening of the current–voltage loop at the origin measures the storage capacitance alone. At positive bias the crossing of the forward and reverse scans then occurs where the storage capacitance and the negative dynamic capacitance cancel. We are analyzing these effects, including the nonzero crossing of current–voltage curves, in separate work. Devices with an internal electromotive force, such as illuminated solar cells, require a more general conduction law, and the dynamic capacitance need not vanish at zero bias.

**Microscopic origin of the delay.** The relaxation time is determined by the transport and accumulation of ions in the active region, such as the salt accumulation near the tip of a conical nanopore or the ionic polarization at a perovskite interface. These processes involve stored species, but they enter the terminal response only through the delayed conductance, as a dynamic capacitance. Connecting $\tau_0$ and $g_a$ to transport models would allow the dynamic capacitance to be predicted from device geometry and material parameters.

**Reliable measurement.** Our results suggest a simple protocol. The impedance and small-step transients at the same bias should give consistent values of $g_a$ and $\tau_0$. The sign change or the vanishing at zero bias should be checked. The amplitude should agree with the stationary current–voltage curve through eq 21. Large pulses from zero bias give the sign and the order of magnitude, and they can be analyzed with eq S1, but additional slow processes must be identified before their areas are interpreted (Figure S3).

**Neuromorphic implications.** Because the inductive response is the electrical signature of potentiation [14], the dynamic capacitance offers a fast and non-destructive impedance test of synaptic functionality. Inductive elements of non-magnetic origin are also the ingredient that produces spiking and oscillation in neuron-like circuits [6, 8]. A quantitative understanding of the dynamic capacitance is therefore a step toward the design of oscillatory neurons and networks built from memristive materials.

## 11. Conclusions

A memristor generates a dynamic capacitance because an internal state controls its conductance. The dynamic capacitance is the time integral of the transient differential conductance after a small voltage step, measured from its final value. It does not measure a stored charge, and it can be much larger than the geometric capacitance. For a single relaxing state it takes the value $-g_a\tau_0$. For a conductance controlled by a state, $I = G(x)\,u$, it vanishes at zero bias, and its sign is set by whether the stationary conductance increases or decreases with the magnitude of the voltage. Potentiation gives a negative capacitance and a chemical inductor; depression gives a positive capacitance with the form of an RC branch. This gives the time-domain criterion of Jonscher and Ershov an explicit formulation, and it accounts for the capacitive and inductive responses of nanofluidic and halide perovskite memristors. In conical nanopore membranes, impedance and voltage pulses give the same sign change and the same order of magnitude of the dynamic capacitance.

## Acknowledgments

The authors acknowledge support from the European Research Council (ERC) via Horizon Europe Advanced Grant, grant agreement nº 101097688 (“PeroSpiker”). Preparation of figures and drafting of the manuscript were assisted by OpenAI ChatGPT/Codex, and Claude (Anthropic).

# Supporting Information

Contents: S1. Impedance fit parameters of the conical multipore membrane. S2. Pulse sequences. S3. Pulse area for large voltage steps. S4. Effect of the slow ionic branch on the transients.

### S1. Impedance Fit Parameters

The impedance spectra of Figure 3a of the main text were fitted with the equivalent circuit of Figure S1a [1]. It contains the series conductance of the solution $g_s$, the membrane capacitance $C_m$, the instantaneous conductance $g_b$ and the relaxing conduction branch $g_a$–$L_a$ of the CALM model, eq 15 of the main text. At high positive bias an additional ionic branch $g_i$–$C_i$ is required. The voltage $u$ is the voltage across the membrane, after the series solution resistance. The relaxing conductance changes sign at zero bias, while $g_b$ remains positive (Figure S1b), and the inductance $L_a = \tau_0/g_a$ carries the sign of $g_a$ (Figure S1d). The membrane capacitance is about 0.2 nF and nearly independent of bias (Figure S1c). The dynamic capacitance shown as circles in Figure 3d of the main text is obtained from these parameters as $C_d = -g_a\tau_0 = -g_a^2 L_a$. For example, at 1 V, $g_a \approx 6$ μS and $L_a \approx 3$ kH give $C_d \approx -100$ nF, and at $-0.4$ V, $g_a \approx -0.4$ μS and $L_a \approx -1.5$ kH give $C_d \approx +0.2$ nF.

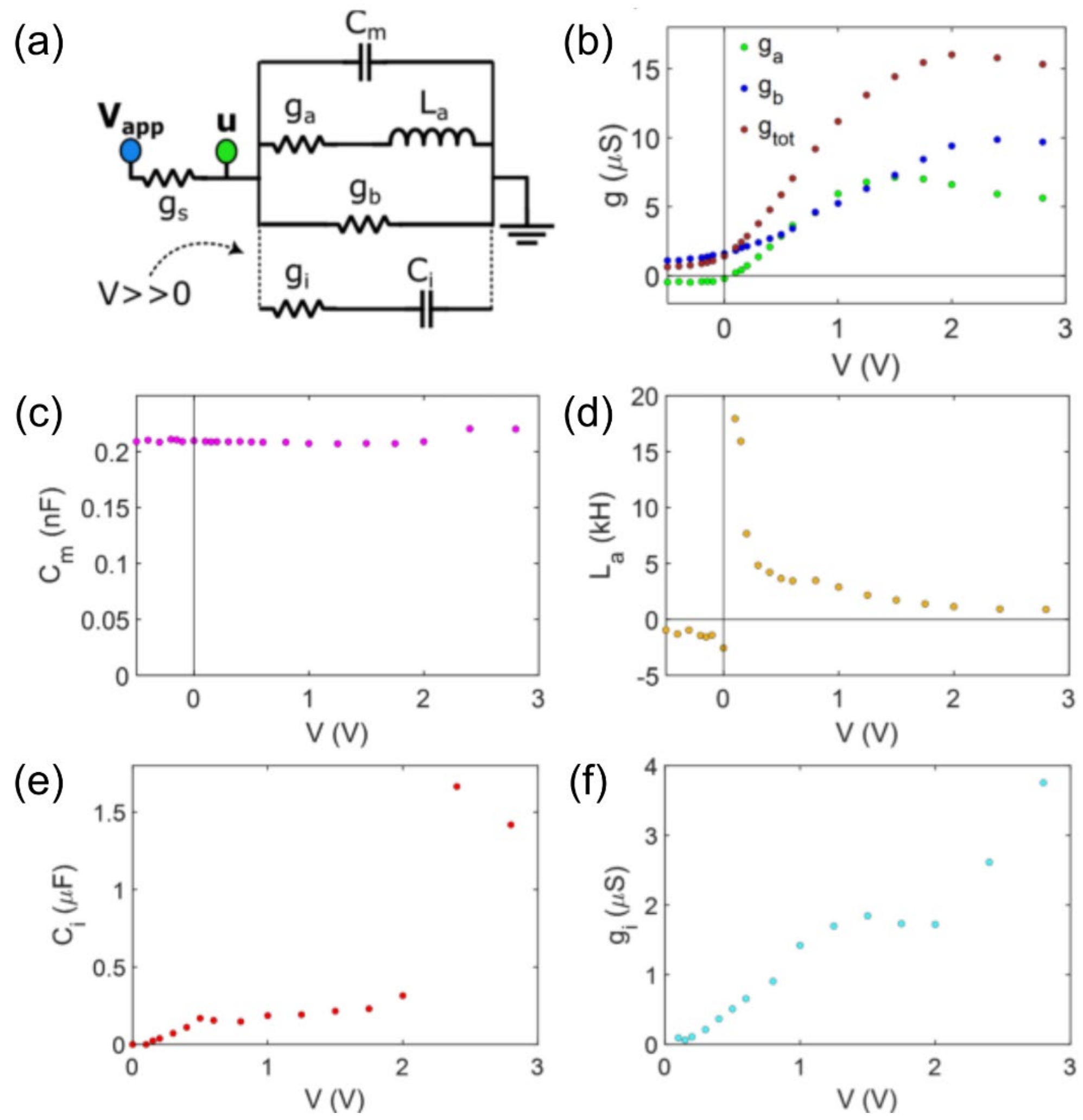

Figure S1. Impedance fit parameters of the conical multipore nanofluidic membrane. (a) Equivalent circuit, with series solution conductance $g_s$, membrane capacitance $C_m$, relaxing conduction branch $g_a$–$L_a$, instantaneous conductance $g_b$, and the additional ionic branch $g_i$–$C_i$ required at high positive bias. (b) Conductances $g_a$, $g_b$ and $g_{\text{tot}} = g_a + g_b$. (c) Membrane capacitance $C_m$. (d) Signed conduction inductance $L_a$. (e) Capacitance $C_i$ and (f) conductance $g_i$ of the additional ionic branch. Adapted from ref 1 under CC BY 4.0.

## S2. Pulse Sequences

The dynamic capacitance shown as stars in Figure 3d of the main text was obtained from sequences of voltage pulses of increasing amplitude, in steps of 0.25 V, applied at positive (Figure S2a,b) and negative (Figure S2c,d) polarity. Each pulse starts from 0 V, and the device returns to 0 V between pulses. For each pulse of amplitude $V$, the current transient was integrated relative to its final value and divided by $V$, as in eq 14 of the main text, with the final value taken at 0.2 s or at the end of the pulse (Figure 3b,c of the main text). Because each pulse is a large step from 0 V, this quantity corresponds to eq S1 (Section S3) rather than to the small-signal value $-g_a\tau_0$ at the bias $V$.

At negative polarity the current is small and nearly constant during each pulse after a short initial decay (Figure S2d), which gives the small positive values of Figure 3d. At positive polarity the current increases during each pulse and the pulse current increases strongly with amplitude (Figure S2b), which reflects the activation of the conductance and gives the large negative values of Figure 3d.

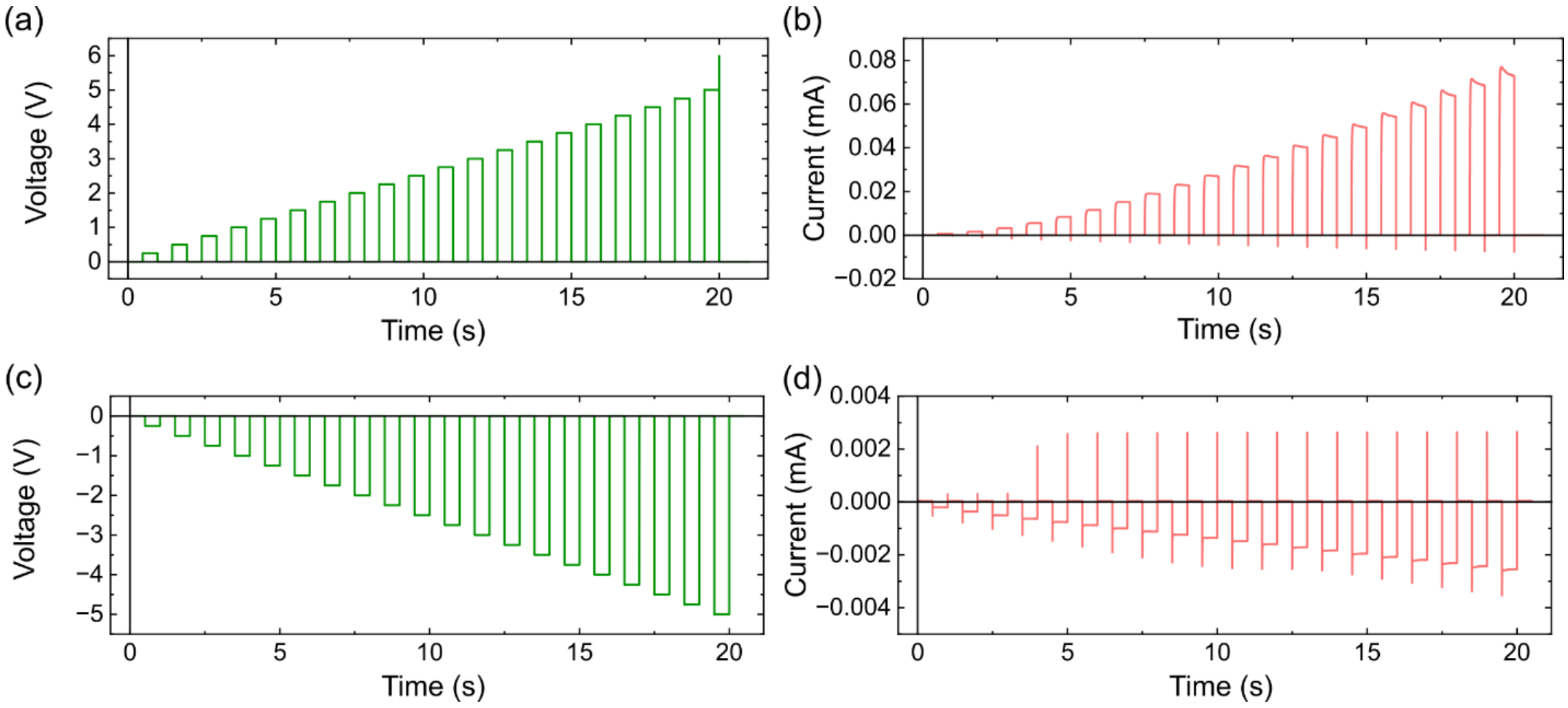

Figure S2. Pulse sequences applied to the conical multipore membrane. (a) Positive voltage pulses of increasing amplitude and (b) the corresponding current. (c) Negative voltage pulses and (d) the corresponding current.

## S3. Pulse Area for Large Voltage Steps

The pulses of Figure S2 are large steps from 0 V, not small perturbations at a bias. For a step from 0 V to $V$ in the CALM model, the voltage is constant after the step, and the state relaxes from $x_{\text{eq}}(0)$ to $x_{\text{eq}}(V)$ with the time $\tau_k(V)$. The area of the conduction transient divided by $V$ is then exactly

$$C_{\text{pulse}} = \frac{1}{V}\int_0^{\infty}[I_c(t) - I_c(\infty)]\,dt = -\tau_k(V)[G_{ss}(V) - G_{ss}(0)] \qquad \text{(S1)}$$

Comparison with eq 21 of the main text shows the origin of the difference. The small-signal value $C_{\text{dyn}} = -g_a\tau_0$ contains the local slope of the stationary conductance, through $g_a$, while the pulse replaces it by the average slope between 0 and $V$. With the parameters of Figure 1 of the main text, eq S1 gives a minimum that is smaller in magnitude than that of $-g_a\tau_0$ ($-69$ nF instead of $-83$ nF), and shifted to higher voltage (1.29 V instead of 1.18 V). It also gives a slower decay at high voltage and larger positive values at negative bias. These are the trends observed in Figure 3d of the main text.

## S4. Effect of the Slow Ionic Branch on the Transients

At high positive voltage the current transient does not approach its final value monotonically. After the initial rise, which is the delayed activation of conduction, the current overshoots and then decays slowly during the rest of the pulse (Figure S3). The slow decay corresponds to the charging of the additional ionic branch of Figure S1a. Its time constant $C_i/g_i$ is about 0.6 s at 2.4 V (Figure S1e,f), comparable to the pulse duration, so that the decay is not complete at the end of the pulse. This branch adds a positive area to the transient and partly compensates the negative area of the conduction relaxation. As a result, the integral depends on whether the final value is taken at 0.2 s or at the end of the pulse (dashed and dotted lines in Figure S3). In the impedance fit the ionic branch is separated from the relaxation of conduction, which is why the impedance gives the more direct estimate of the dynamic capacitance at high voltage.

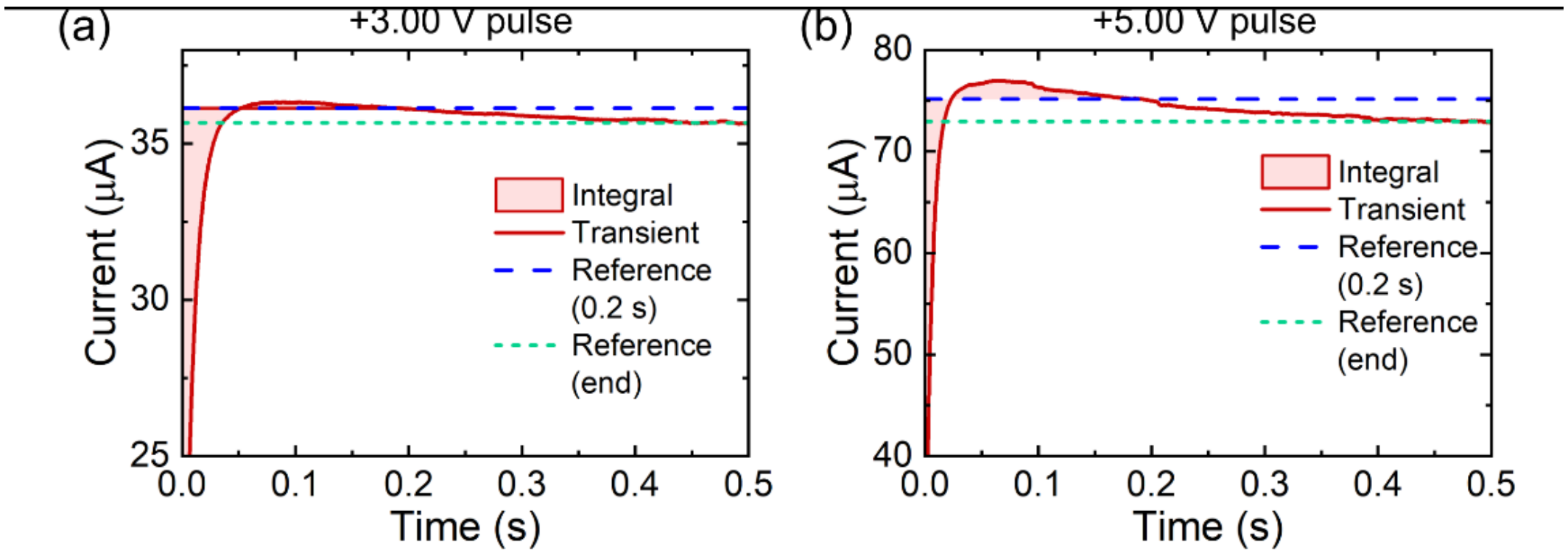

Figure S3. Current transients during pulses of (a) +3.00 V and (b) +5.00 V. The shaded areas are the transient integrals. The dashed and dotted lines mark the final value taken at 0.2 s and at the end of the pulse, respectively.